\documentclass[english,superscriptaddress,twocolumn]{revtex4-1}
\usepackage[LGR,T1]{fontenc}
\usepackage[latin9]{inputenc}
\usepackage{babel}
\usepackage{textcomp}
\usepackage{amstext}
\usepackage{graphicx}
\usepackage{subscript}
\usepackage[unicode=true,pdfusetitle,
 bookmarks=true,bookmarksnumbered=false,bookmarksopen=false,
 breaklinks=false,pdfborder={0 0 1},backref=false,colorlinks=false]
 {hyperref}

\makeatletter

\DeclareRobustCommand{\greektext}{%
  \fontencoding{LGR}\selectfont\def\encodingdefault{LGR}}
\DeclareRobustCommand{\textgreek}[1]{\leavevmode{\greektext #1}}
\DeclareFontEncoding{LGR}{}{}
\DeclareTextSymbol{\~}{LGR}{126}

\makeatother

\begin{document}

\title{Out of equilibrium anomalous elastic response of a water nano-meniscus}

\author{Simon Carpentier}

\affiliation{Univ. Grenoble Alpes, F-38000 Grenoble, France }

\affiliation{CNRS, Inst NEEL, F-38042 Grenoble, France}

\author{Mario S. Rodrigues}

\author{Miguel V.Vitorino }

\affiliation{Uniservity of Lisboa, Falculty of Sciences, BioISI-Biosystems \&
Integrative Sciences Institute, Campo Grande, Lisboa, Portugal }

\author{Luca Costa}

\affiliation{ESRF, The European Synchrotron, 71 Rue des Martyrs, 38000 Grenoble,
France}

\author{Elisabeth Charlaix}

\affiliation{Univ. Grenoble Alpes, F-38000 Grenoble, France }

\affiliation{CNRS, LIPhy, Grensoble, F-38402, France }

\author{Joël Chevrier}

\affiliation{Univ. Grenoble Alpes, F-38000 Grenoble, France }

\affiliation{CNRS, Inst NEEL, F-38042 Grenoble, France}
\begin{abstract}
We report the observation of a transition in the dynamical properties
of water nano-menicus which dramatically change when probed at different
time scales. Using a AFM mode that we name Force Feedback Microscopy,
we observe this change in the simultaneous measurements, at different
frequencies, of the stiffness G\textquoteright (N/m), the dissipative
coefficient G\textquotedblright (kg/sec) together with the static
force. At low frequency we observe a negative stiffness as expected
for capillary forces. As the measuring time approaches the microsecond,
the dynamic response exhibits a transition toward a very large positive
stiffness. When evaporation and condensation gradually lose efficiency,
the contact line progressively becomes immobile. This transition is
essentially controlled by variations of Laplace pressure. 
\end{abstract}
\maketitle
Visco-elastic properties of water nanobridges\cite{thomson18724}
at very different time scales, have never been investigated despite
ubiquitous presence of capillarity. Associated forces are among the
most intense at nanoscales with important consequences in soils and
granular media. Interest in dynamical properties is immediately raised
if one considers interacting surfaces with roughness scales down to
nanometer. Even at moderate speeds, such as v=1m/s, characteristic
times of surface interaction down to microsecond appear in these conditions.
Our measurements approaching these time scales, further strengthen
the relevance of the dynamical properties to describe how real surfaces
interact and are certainly of crucial importance in numerous AFM experiments
\cite{riedo2002kinetics}. We here report measurements of dynamical
properties of a water nanobridge for a continuous range of the surface
gap and a frequency bandwidth up to 0.1 MHz. We identify two regimes:
one is the thermodynamical equilibrium; the second is out of equilibrium.
Evaporation and condensation of water molecules between the liquid
and the gas phase ensures that the nano-meniscus curvature is the
one at thermodynamical equilibrium(2H=-1/r\textsubscript{k}) where
2H is the water bridge curvature and r\textsubscript{k} is the Kelvin
radius. At time short enough, molecule exchanges between the liquid
and the gas phase are no longer efficient and the water nanobridge
is led to acquire a constant volume. The liquid bridge relaxation
time is the time needed for the bridge to adapt its shape as required
by thermodynamical equilibrium, when its length h is abruptly changed
by \textgreek{d}h. This is controlled by molecular transport through
diffusion mechanisms in gas phase. This relaxation time \textgreek{t}
can be estimated as in SFA context, see Ref.\cite{crassous1995etude}. 

\begin{equation}
\tau=2\gamma\rho r{{}^2}ln(R/\rho)/P_{sat}\,r_{k}{{}^2}\,D\label{eq:1}
\end{equation}

D is the diffusion coefficient of water molecules in air (D= 0.282
cm\textsuperscript{2}/sec from Ref.\cite{cussler2009diffusion}).
\textgreek{r} is the bridge azimuthal radius, r, its meridional radius.
P\textsubscript{sat} is the saturated vapor pressure, R is the typical
scale of the overall system and \textgreek{g} the surface tension
of the vapor liquid interface. In our experimental conditions, with
a Kelvin radius r\textsubscript{k} of about 12nm, the characteristic
time is found to be around \textgreek{t}=10\textsuperscript{-6} second.
As \textgreek{w}/2\textgreek{p} increases from 300Hz to 114kHz, in
our experiment, \textgreek{wt} evolves from 2.10\textsuperscript{-3}
to 0.7. At \textgreek{wt} 2.10\textsuperscript{-3}, the liquid nanobridge
should be observed at thermodynamical equilibrium. At \textgreek{wt}=0.7,
\textgreek{w}/2\textgreek{p}=114kHz, this time \textgreek{t} is too
short for the thermodynamical equilibrium to settle during oscillation.
We shall therefore consider an extreme regime with constant volume
when dealing with analysis of experiments performed at these high
frequencies. 

\begin{figure}
\includegraphics[width=9cm]{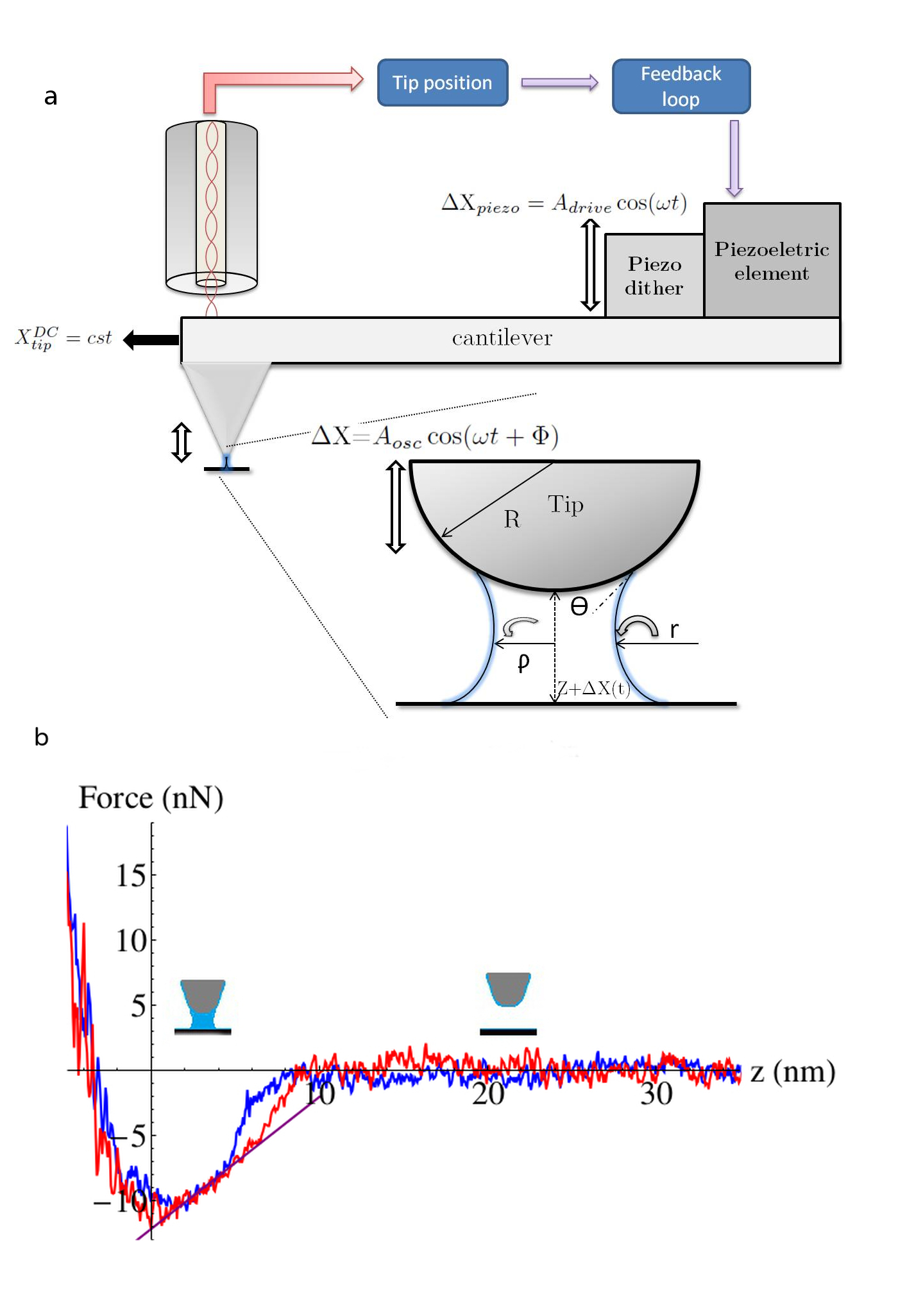}

\protect\caption{(a) Description of the FFM measurement strategy. The DC tip position
is a constant. This is due to real time application of a feedback
force determined through a control loop after measurement of the tip
position. Piezodither enforces an AC tip oscillation with 1 nm of
free oscillation amplitude. A\protect\textsubscript{osc} and \textgreek{F}
are the measured quantities. (b), Static force measured between a
brand new silicon AFM tip and a plane silicon sample during approach
(blue) and retract (red) by FFM in presence of a capillary bridge.
The linear behavior is fitted using the approximate solution given
by J.Crassous et al. in Ref.\cite{Crassous2011} with R=14nm and r\protect\textsubscript{k}=12nm.\label{fig:1}}
\end{figure}

Using the AFM mode that we name Force Feedback Microscopy (FFM), we
have measured simultaneously the static force, the interaction stiffness
G' and the associated dissipation G'' at different frequencies. As
described in Fig.\ref{fig:1}.(a), the FFM used to perform this experiment
is based on a homemade AFM setup with optical fiber \cite{rugar1989improved}.
These three simultaneous measurements are done using a single nanotip-microlever
system, with excitation frequency varied from 300Hz up to 114kHz.
During FFM experiments, a feedback force is applied in real time to
the tip so that it cancels the force due to the surface (this suppresses
the \textquotedblleft jump to contact\textquotedblright{} observed
in classical AFM force-approach curve). The total force applied to
the tip is then equal to zero and the tip remains immobile. To apply
this force in real time, a piezoelement simply changes the DC position
of the clamped end part of the microlever. This measured displacement
multiplied by the lever stiffness k results in the static tip/surface
force measurement. Measured on this basis \cite{rodrigues2012atomic,costa2013comparison},
the static force is reported in Fig.\ref{fig:1}.(b). Both the tip
radius and the Kelvin radius, r\textsubscript{k} are then estimated.
 The minimum force is roughly -4\textgreek{pg}R, which leads to R
close to 14nm, see more details in Ref.\cite{Crassous2011}. The slope
of the force versus distance as the AFM tip is retracted, gives r\textsubscript{k}=12nm
\cite{kohonen2000capillary,inset1}. As shown in Fig.\ref{fig:1}.(a),
a nanometeric oscillation is imposed to the tip. Measure of the tip
amplitude and the associated phase shift leads to the determination
of the stiffness interaction and the associated dissipation following
these linear equations (see details in Ref.\cite{rodrigues2014system}):
\begin{equation}
G'=F_{r}[ncos(\Phi)-cos(\Phi_{\infty})]\label{eq:G' ffm}
\end{equation}

\begin{equation}
G''=\frac{F_{r}}{\omega}[\text{\textminus}nsin(\text{\textgreek{F}})+sin(\text{\textgreek{F}}_{\infty})]\label{eq:G'' ffm}
\end{equation}

\textgreek{F} is the phase shift between excitation at the lever clamped
end and the tip oscillation. n is the normalized amplitude of tip
vibration i.e. the measured amplitude at distance z divided by the
measured amplitude when tip is far away from the surface. At large
distance n=1. In the experimental environment used (normal conditions
in air), values of F\textsubscript{r} and \textgreek{F}\textsubscript{\ensuremath{\infty}}
are directly obtained from the system transfer function when the tip
is far from the surface.

\begin{equation}
F_{r}=[(k-m\omega^{2})^{2}+c^{2}\omega^{2}]^{1/2}\label{eq:Fr}
\end{equation}

\begin{equation}
\text{\textgreek{F}}_{\infty}=arctan[(c/m)\omega/(\omega^{2}\text{\textendash}\omega{}_{0}^{2})]\label{eq:phy infiny}
\end{equation}

\begin{figure}

\includegraphics[width=9cm]{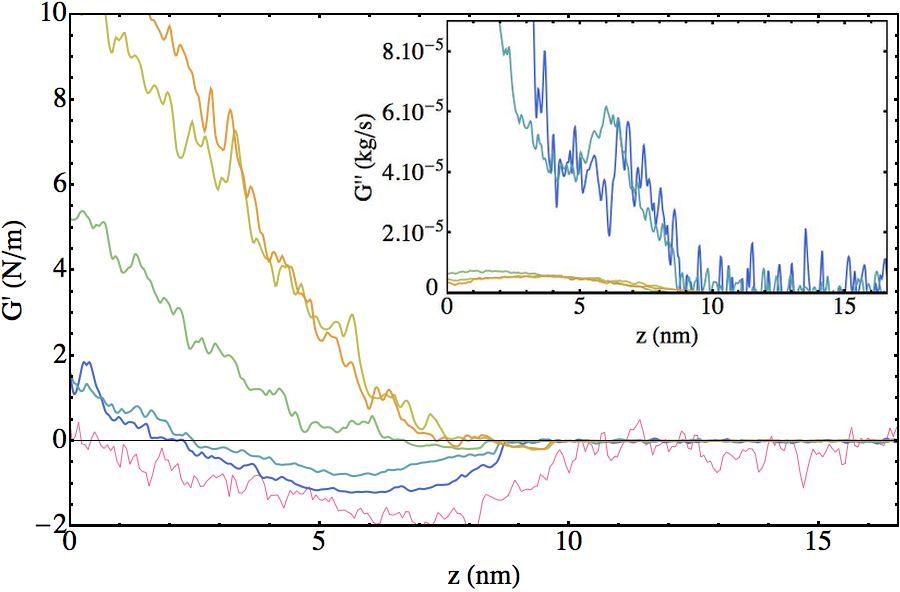}\protect\caption{From blue to orange curves (respectively 300 Hz, 1 kHz, 44 kHz, 94
kHz and 114 kHz) show interaction stiffness, G\textquoteright{} versus
tip surface distance as deduced from experimental measurements using
Eq.\ref{eq:G' ffm}. Red curve is the negative of the numerical derivative
of the measured static force as the tip is pulled away from the surface.
Inset: G\textquotedblright (kg/sec) result from Eq.\ref{eq:G'' ffm}.
It characterizes the mechanical energy dissipation increase during
tip oscillation due to the capillary bridge. G\textquoteright \textquoteright{}
at high frequencies is distinctively decreased as compared to G\textquotedblright{}
at low frequencies. Dissipation when in the thermodynamical equilibrium
is clearly much higher. \label{fig:figure 2}}

\end{figure}

In the used frequency range with first resonance at 74,330kHz, a single
mode description of the lever dynamics is sufficient. The stiffness
of the lever k is obtained from measurement of the Brownian motion.
The damping coefficient c and \textgreek{w}\textsubscript{0} (and
therefore m) are obtained from lever transfer function measurement
when the tip is far from the surface. F\textsubscript{r} and \textgreek{F}\textsubscript{\ensuremath{\infty}}
determined using this method are inserted in Eq.\ref{eq:G' ffm} and
\ref{eq:G'' ffm} to obtain G\textquoteright{} and G\textquoteright \textquoteright{}
from n and \textgreek{F}. In Fig.\ref{fig:figure 2}, k=2 N/m, c=2.49
10\textsuperscript{-8} kg/sec and \textgreek{w}\textsubscript{0}/2\textgreek{p}=74,330kHz. 

Fig.\ref{fig:figure 2} shows that the visco-elastic coefficients
exhibit a strong variation as the excitation frequency is varied over
close to 3 orders of magnitude: G\textquoteright{} varies from -1N/m
up to 10N/m (therefore from negative to positive). The positive stiffness
measured at high frequency, G\textquoteright =+10N/m is a very high
interaction stiffness in AFM field. It is surprisingly observed associated
to water and to an intense attractive static force close to 10nN.
From this apparent positive stiffness, although only surface effects
due to the nano-meniscus are at work, an effective Young modulus E
can be formally deduced: E = G\textquoteright z/\textgreek{p} r\textsubscript{b}.
It is in the range 0.1-1 GigaPa and increases as the tip moves closer
to the surface. 

At \textgreek{w}/2\textgreek{p}=300Hz, the measured interaction stiffness
G\textquoteright{} corresponds to the numerical derivative of the
static force versus distance (i.e. the derivative of the red curve
in Fig.\ref{fig:1}(b)). Here \textgreek{wt}<\textcompwordmark{}<1
and the water nanobridge appears in thermodynamical equilibrium. However,
the gradual increase of the excitation frequency leads to an important
increase of the interaction stiffness, which finally becomes positive
at all investigated distances \cite{inset2}. The high frequency regime
with positive and high stiffness is therefore an out of equilibrium
behavior of the water droplet. As determined by use of Eq. \ref{eq:1},
an estimate of the bridge relaxation time \textgreek{t} is 10\textsuperscript{-6}sec:
the water molecule exchange between liquid and gas diminishes as the
frequency increases toward 1MHz. This ultimately leads to a regime
with a water nanobridge that has a constant volume. With this hypothesis
of constant volume at high frequency, we analyze the observed positive
stiffness G' due to the water nano-menicus properties. This in fact
follows analysis of Lambert et al in Ref.\cite{valsamis2013vertical}
produced for far larger water bridge (millimeter range). Equation
\ref{eq:F} shows the expression of the total force due to the water
nano-menicus on the tip:

\begin{equation}
F=2\pi r_{b}\gamma\,\sin(\theta)-2H\pi r_{b}^{2}\gamma\label{eq:F}
\end{equation}

\begin{figure}
\includegraphics[width=9cm]{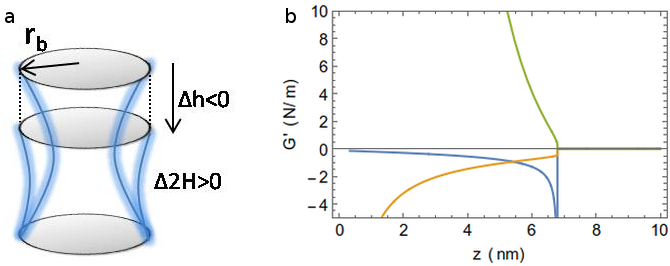}

\protect\caption{(a), evolution of the capillary bridge shape as the gap decreases
at constant volume and at locked contact line. (b), results from numerical
calculation, reproduce the key experimental features obtained using
Force Feedback Microscopy at different frequencies. At low frequency
(blue curve), the signature of the thermodynamical equilibrium is
again obtained with a negative stiffness at all distances. At high
frequency oscillation (green curve), the water nanobridge is at constant
volume and the contact line is immobile. A positive stiffness is obtained
like in experimental results. It is however significantly above all
measured stiffnesses for which, at \textgreek{wt}=0.7, contact line
mobility is reduced but certainly far from zero. The orange curve
represents a regime at constant volume, but with a contact line free
to move. It leads to a negative stiffness. This regime, which implies
a curvature only marginally varying, cannot explain the experimental
observations at high frequency. \label{fig:3}}
\end{figure}

The force decomposes in two terms. The first term of Eq.\ref{eq:F}
is due to the tension associated to the contact line whereas the second
is due to the Laplace pressure. \textgreek{g} is the water surface
tension, 2\textgreek{p}r\textsubscript{b} the length of the circular
contact line (or triple line), (2H=1/\textgreek{r}-1/r) and \textgreek{j},
the angle between the water bridge and the surfaces. At thermodynamic
equilibrium (\textgreek{wt}<\textcompwordmark{}<1) with efficient
evaporation condensation processes, whatever the tip surface distance
and the oscillation amplitude, H is a negative constant. The pressure
inside the bridge then does not change, and the contact line moves
to accommodate this constrain of constant curvature as the tip surface
distance is varied. During shortening of the bridge (\textgreek{d}h<0),
the radius r\textsubscript{b} increases as a result of the bridge
spreading: the resulting stiffness \textgreek{d}F/\textgreek{d}h is
negative as observed experimentally. At \textgreek{wt}>\textcompwordmark{}>1,
we consider that the nanobridge volume remains constant. If the contact
line does not move, the radius r\textsubscript{b} is constant and
the curvature increases (H<0) as described in Fig.\ref{fig:3}(a).
The second term in the force then provides a contribution to stiffness
that is:

\begin{equation}
\frac{\delta F}{\delta h}=-2\pi r_{b}^{2}\gamma\frac{\delta H}{\delta h}\label{eq:deltaF/deltah}
\end{equation}

This \textgreek{d}F/\textgreek{d}h is then a positive contribution,
which explains the origin of the positive stiffness experimentally
observed.

A numerical calculation of the stiffness G' in the three extreme regimes
(thermodynamical equilibrium, constant volume with locked contact
line or free to move contact line) is reported in Fig.\ref{fig:3}(b). 

Finally as shown in inset of Fig.\ref{fig:figure 2}, G\textquotedblright (kg/sec),
the dissipative part of the linear response, is found to decrease
as the excitation frequency increases and as G\textquoteright{} severely
increases. This leads us to conclude that this change of G\textquotedblright{}
is also determined by the contact line behavior. A mobile contact
line on the AFM tip then appears to be associated to an important
dissipation. As mobility of the contact line decreases, this channel
of energy dissipation closes and G\textquotedblright{} diminishes
accordingly. Although only surface effects are here at work, G\textquotedblright{}
can be formally analyzed as an apparent volume viscosity: G\textquotedblright \ensuremath{\propto}\textgreek{h}R,
where R=14nm is the tip size. In that case, with a typical G\textquotedblright =10\textsuperscript{-5}
kg/sec here measured and taken equal to \textgreek{h}R, \textgreek{h}
is in the range 10\textsuperscript{2}-10\textsuperscript{3}Pa.sec.
This is orders of magnitude above \textgreek{h}= 10\textsuperscript{-3}Pa.sec,
the bulk value of water viscosity(Ref.\cite{Inset3,khan2010dynamic}).
\\

In conclusion, for the first time, two different visco-elastic regimes
in dynamical properties of a water nanobridge are measured. The major
result is that, despite the fact that the static interaction between
surfaces associated to capillary bridges is always attractive, an
apparent and highly positive stiffness appears at the nanoscale when
the two surfaces interact with a short characteristic time, whereas,
the friction measured in same conditions, is found strongly reduced.
A description can be proposed to simultaneously explain these results.
Contrary to the situation at thermodynamical equilibrium with condensation
and evaporation, in the out of equilibrium regime, any displacement
of the line can only be ensured by a flow in the nanodroplet. The
sticky boundary condition leads to a diverging shear velocity as the
water film thickness goes to zero close to the triple line. In this
situation, moving the line becomes extremely difficult. Indeed, we
observe in the diminution of G\textquotedblright , at high frequencies
a drop of the line mobility. In this case, at constant volume, there
is then no alternative: the curvature of the nano-meniscus must change,
which we observe in the increase of G\textquoteright . 

The frequency ranges in our experiments are the ones used to operate
AFM in dynamical mode. Results presented here will certainly have
to be considered in the daily analysis of dynamic mode AFM experiences
when a water nanobridge is present. Due the importance of the observed
changes in effective visco-elastic properties of water at nanoscale,
this conclusion is certainly not limited to AFM but opens a broader
perspective, as capillary bridges are now known to be important in
surface interactions not only at different scales but also at different
time scales. 
\begin{acknowledgments}
Mario S. Rodrigues and Miguel V. Vitorino acknowledge financial support
from Fundação para a Ciência e Tecnologia, grants SFRH/BPD/69201/2010
and PD/BD/105975/2014 respectively. 
\end{acknowledgments}

\bibliographystyle{unsrt}
\bibliography{librarynanoletters}

\begin{thebibliography}{10}

\bibitem{thomson18724}
William Thomson.
\newblock 4. on the equilibrium of vapour at a curved surface of liquid.
\newblock {\em Proceedings of the Royal Society of Edinburgh}, 7:63--68, 1872.

\bibitem{riedo2002kinetics}
Elisa Riedo, Francis L{\'e}vy, and Harald Brune.
\newblock Kinetics of capillary condensation in nanoscopic sliding friction.
\newblock {\em Physical review letters}, 88(18):185505, 2002.

\bibitem{crassous1995etude}
J{\'e}r{\^o}me Crassous.
\newblock {\em Etude d'un pont liquide de courbure nanom{\'e}trique:
  propri{\'e}t{\'e}s statiques et dynamiques}.
\newblock PhD thesis, 1995.

\bibitem{cussler2009diffusion}
Edward~Lansing Cussler.
\newblock {\em Diffusion: mass transfer in fluid systems}.
\newblock Cambridge university press, 2009.

\bibitem{rugar1989improved}
D~Rugar, HJ~Mamin, and Peter Guethner.
\newblock Improved fiber-optic interferometer for atomic force microscopy.
\newblock {\em Applied Physics Letters}, 55(25):2588--2590, 1989.

\bibitem{rodrigues2012atomic}
Mario~S Rodrigues, Luca Costa, Jo\"{e}l Chevrier, and Fabio Comin.
\newblock {Why do atomic force microscopy force curves still exhibit jump to
  contact?}
\newblock {\em Applied Physics Letters}, 101(20):203105, 2012.

\bibitem{costa2013comparison}
Luca Costa, Mario~S Rodrigues, Simon Carpentier, Pieter~Jan van Zwol, Joël
  Chevrier, and Fabio Comin.
\newblock Comparison between atomic force microscopy and force feedback
  microscopy static force curves.
\newblock {\em arXiv preprint arXiv:1306.2775}, 2013.

\bibitem{Crassous2011}
J\'{e}r\^{o}me Crassous, Matteo Ciccotti, and Elisabeth Charlaix.
\newblock {Capillary force between wetted nanometric contacts and its
  application to atomic force microscopy.}
\newblock {\em Langmuir : the ACS journal of surfaces and colloids},
  27(7):3468--73, April 2011.

\bibitem{kohonen2000capillary}
Mika~M Kohonen and Hugo~K Christenson.
\newblock Capillary condensation of water between rinsed mica surfaces.
\newblock {\em Langmuir}, 16(18):7285--7288, 2000.

\bibitem{inset1}
This measured value of rk is higher than expected based on vapor pressure. We
  have not controlled the hygrometry level during numerous experiments. We
  estimate an average value of hygrometry of 50 per cent. This would lead to a
  Kelvin radius rk clearly smaller than 5nm. A hygrometry level close to 99 per
  cent is required to observe a Kelvin radius of about 12nm. This is certainly
  not our case. It is however known that with no special treatment of the used
  silicon surfaces, large Kelvin radii are commonly observed, see previous
  reference. rk=12nm is the value we have used to determine the characteristic
  time.

\bibitem{rodrigues2014system}
Mario~S Rodrigues, Luca Costa, Jo{\"e}l Chevrier, and Fabio Comin.
\newblock System analysis of force feedback microscopy.
\newblock {\em Journal of Applied Physics}, 115(5):054309, 2014.

\bibitem{inset2}
At w=44kHz and above, the measured stiffness reaches values between 6N/m and
  10N/m. Direct contact between the tip and a hard surface with Pauli repulsion
  at work, could be responsible for such a high stiffness. All curves presented
  in Fig. 2 have been measured during pulling the tip away from the surface. At
  all frequencies, we used exactly the same experimental conditions: tip
  control, lever oscillation amplitude, and overall time to acquire data along
  a complete approach/retract force curve. We simultaneously measured the
  static force and the dynamic response, so that we know at each point, what is
  the DC tip position in the static force curve shown in Fig. 1(b). The same
  conditions, at identical tip surface distance, with comparable oscillation
  amplitude, lead to a negative stiffness at low frequencies and to a positive
  one at high frequency. We then conclude that this increase is a direct
  consequence of the water bridge dynamical properties and that at high
  frequencies, the water bridge becomes much stiffer.

\bibitem{valsamis2013vertical}
J-B Valsamis, Massimo Mastrangeli, and Pierre Lambert.
\newblock Vertical excitation of axisymmetric liquid bridges.
\newblock {\em European Journal of Mechanics-B/Fluids}, 38:47--57, 2013.

\bibitem{Inset3}
Using small-amplitude AFM based technique, variations of dynamical properties
  of a confined liquid such as a comparable increase of the stiffness, have
  been reported in Ref.15. This leads authors to conclude that bulk properties
  of water are changed by confinement close to a surface. These reports are
  related to change in molecular structure of liquid layers whereas in our
  case, the nano-meniscus behavior is the key. In Ref.15, the tip is immersed
  into the bulk liquid. There is no nano-meniscus due to a gas-liquid
  interface. The strong reported variations occur when the tip-surface distance
  is below 1nm, as in our case the relevant separations are between 1nm and
  10nm.

\bibitem{khan2010dynamic}
Shah~H Khan, George Matei, Shivprasad Patil, and Peter~M Hoffmann.
\newblock Dynamic solidification in nanoconfined water films.
\newblock {\em Physical review letters}, 105(10):106101, 2010.

\end{thebibliography}

\end{document}